# Deep Learning for Size and Microscope Feature Extraction and Classification in Oral Cancer: Enhanced Convolution Neural Network

Prakrit Joshi[1], Omar Hisham Alsadoon[2,] Abeer Alsadoon[1,3,4,5*], Nada AlSallami[6] , Tarik A. Rashid[7], P.W.C. Prasad[8], Sami Haddad[9]

[1] School of Computing Mathematics and Engineering, Charles Sturt University (CSU), Australia
[2] Department of Islamic Sciences, Al Iraqia University, Baghdad, Iraq
[3] School of Computer Data and Mathematical Sciences, Western Sydney University (WSU), Sydney, Australia
[4] Kent Institute Australia, Sydney, Australia
[5] Asia Pacific International College (APIC), Sydney, Australia
[6] Computer Science Department, Worcester State University, MA, USA
[7] Computer Science and Engineering, University of Kurdistan Hewler, Erbil, KRG, IRAQ
[8] Department of Oral and Maxillofacial Services, Greater Western Sydney Area Health Services, Australia

**Abeer Alsadoon[1*]**

* Corresponding author. A/Prof Abeer Alsadoon, [1] School of Computing Mathematics and Engineering, Charles Sturt University (CSU), Sydney Campus, Australia, Email: alsadoon.abeer@gmail.com , Phone +61 2 9291 9387

## Abstract

*Background and Aim:* Over-fitting issue has been the reason behind deep learning technology not being successfully implemented in oral cancer images classification. The aims of this research were reducing overfitting for accurately producing the required dimension reduction feature map through Deep Learning algorithm using Convolutional Neural Network. *Methodology:* The proposed system consists of Enhanced Convolutional Neural Network that uses an autoencoder technique to increase the efficiency of the feature extraction process and compresses information. In this technique, unpooling and deconvolution is done to generate the input data to minimize the difference between input and output data. Moreover, it extracts characteristic features from the input data set to regenerate input data from those features by learning a network to reduce overfitting. *Results:* Different accuracy and processing time value is achieved while using different sample image group of Confocal Laser Endomicroscopy (CLE) images. The results showed that the proposed solution is better than the current system. Moreover, the proposed system has improved the classification accuracy by 5~ 5.5% on average and reduced the average processing time by 20 ~ 30 milliseconds. *Conclusion:* The proposed system focuses on the accurate classification of oral cancer cells of different anatomical locations from the CLE images. Finally, this study enhances the accuracy and processing time using the autoencoder method that solves the overfitting problem.

## 1. Introduction

In conventional practice, a pathologist investigates the histopathological images of oral mucosa, i.e. the study of tissue samples of the affected area by a microscope. The pathologist visually examines the image, and the keratinized area and this manual assessment process fully depend on the pathologist's expertise and experience [1]. But, this manual process is time-consuming and prone to diagnostic errors, which can be solved by using deep learning technology which automates the image classification of cancer cells. It provides an accurate classification of extensive image dataset by training the algorithm with the experts' knowledge [2].

Deep learning has been applied to several applications, particularly in the field of medicine for medical image analysis [3]. Deep learning provides an advanced classification technology enabling calculation models consisting of several processing layers to learn the data representations used for oral cancer image classification. The algorithm is trained using expert knowledge, and the trained network is tested using the remaining dataset, which provides the exact feature map that results in highly accurate image classification [2]. The network arrangement and lack of proper training dataset results in the network not precisely producing the required feature map with dimension reduction, which could overfit the network. The solution to reduce this problem is using a convolutional neural network classifier. This classifier is further modified to generate the exact feature map from the input dataset learning required for accurate oral cancer classification.

A current study of Deep Learning uses various algorithms and techniques for better feature extraction and image classification. For accurately classifying oral cancer images, it is necessary to train the algorithm using the proper set of the input dataset with a less complex neural network. The current state of art using Inception v3 convolutional neural network provides classification accuracy of 87.02% [4]. Research on this network has identified overfitting problem which cannot produce the exact required feature map of classification accuracy. In order to overcome this [5], introduced autoencoder architecture for extracting characteristic features from the input data but still suffers from reconstruction error. Therefore, current studies still have an area that needs be improved.

The contribution o of this paper summarised in the below points

- Increase the classification accuracy and speed for image classification of oral cancer.
- This research aims to provide highly accurate feature map, input dimension reduction and high classification accuracy by decreasing overfitting.
- This study proposes an autoencoder architecture that extracts characteristic features from the input dataset, which regenerates the input data from those features by learning a network [5]. In addition, the input is passed for de-convolution and de-pooling to extract the optimized image input. This helps extract more complex image features with higher-order structures.

## 2. Literature Review

The main aim of this review is to conduct a survey of different available papers and find the new solution to improve the existing system. The following section provides a review of different papers and provides an idea about the existing solution to the problem. [2] modified the pre-trained Convolutional Neural Network, the Google Net Inception V3 CNN architecture by using the regression-based partition convolution and subsampling layer to use optimum data into the training network. The author improved the state of art's solution [6]. Their structure processes complex data which produced classification accuracy of 94.5% with specificity 0.98 and sensitivity of 0.94, which is greater than another base classifier in a single phase of training data set. But the classifier has not been designed to extract the feature map of fewer variation data which need to be considered to increase the classification accuracy.

[7] employed 50 CBCT 3D image dataset verified by the expert for identifying periapical cyst and keratocystic odontogenic tumor (KCOT) lesions for exact classification. They offered the solution to the problem performing segmentation on Cone Beam Computed Tomography (CBCT) images using the viewer software. The author marked the lesional volume of interest and calculated order statistics for each CBCT dataset which was not marked in the previous[8] solution. This solution identifies the best classifier that is SVM with the accuracy percentage of 96% and F1 score of 96%, which is the best amongst other classifiers. The classifier performance has been increased by decreasing the size of the feature vector. BCT dataset needs to be enhanced with different types of dental pathologies to enhance accuracy. [4] studied a different classification approach for classifying Confocal Laser

Endomicroscopy (CLE) images and considered patch probability fusion method to outperform other conventional approaches. The author improved the state of art's solution [9]. In this method, the author designed the network with LeNet-5 network, 2 convolutional layers with different filter sizes with max-pooling layer, only fully connected layer and output softmax output. This design has increased the classification accuracy to 87.02% at a sensitivity of 90.71% and specificity of 83.80%. However, the classification can be improved further by adding entities like precursor affected region of cancer and adding histopathology process to assess the tissue.

[5] experimented using a different network architecture and extended the network to process larger pathological input images to evaluate the local phenotypic feature and their distribution in the tissue. This solution has improved the current state of art solution [10]. They offered a solution to implement deep convolutional autoencoders to extract the characteristic features from the given inputs. Further, they built a sample based on three autoencoders and one classification reducer to evaluate larger pathological images classifying the transcriptome subtypes. The complex feature extraction increased with the larger input image size, and hence the accuracy of classification also increased to 98.89%. However, it is hard to differentiate the statistical distribution of cellular features in bigger tissue input. Thus, the latest approach can be developed for the differentiation of various tissue types. [11] modified the AlexNet architecture to dropout 1 layer from the fully connected layer and used 5 layers as convolutional layers and 2 as fully connected layers and used dropout probability of 0.5 to make the classification appropriate for the small-size data analysis which is the improved solution [12]. It helped to quickly analyze the minimum number of clinical images that are ready for the training of dataset. This resulted in an accuracy of 78.2% which result in matches to that of the 2 experienced radiologists study results. While the diagnostic performance of the deep learning system had an accuracy of 78.2% on using the arbitrarily sized square image patches as an input, there was no consideration of impact of datasets and input image size sampling on the training. But this model failed to operate in real time due to manual image segmentation.

[13] has added the detection operation in segmentation operation, which alone focuses on a particular reason containing an organ and segments this particular organ from others. This solution has improved the current state of the art solution [14]. This solution has reduced the overall detection time to 16s and segmentation time to 30s for labelling of a single image input which is suitable for the clinical workflow and is more optimized than other solutions such as Fully Convolutional Neural Network (FCN). Although the solution has improved the sensitivity from 0.997 to 1 for the most organs, the input is only taken from the non-contrast CT scanning. Which limited the accuracy of detection and segmentation. Thus, the current Organs-at-risk detection and segmentation network (ODS net)) can be improved by considering both the non-contrast and contrast CT images. [15] have fused the autofluorescence and white light images information into three different image channel and feed the information into the deep learning neural network. This solution has improved the current state of art solution [16]. The small size network with five different convolutional networks is used to train the small data sets to reduce the network complexity. Hence, this system provides the average accuracy of 86.9% with sensitivity of 85% and specificity of 88.7% when applied in a 4-fold cross validation. Which improved the performance in detecting the oral cancer and provided the cost effective solution to many smartphone users for further test and treatment.

[17] used transfer learning by using the greater sample dataset on the radiographic dataset with known biopsy for preparing training input data set. They prepared test data using 50 ameloblastoma images and 50 Keratocystic Odontogenic Tumor (KCOT) images to overcome the limitation of minimum training dataset. The author pre-trained the VGG-16 (16-layer CNN) network in ImageNet and developed the gradient weighted class activation maps (Grad-CAM) to identify the unequal regions on panoramic digital X-ray images. This solution has improved the current state of art solution [18]. It provides an accuracy of 83.0 % and time to analyze the images is 38 seconds as compared to 23.1 minutes time taken by the oral and maxillofacial surgeons for classification [18]. However, the lateral X-ray radiographs have not been considered for image input. Both frontal X-ray and lateral view X-rays should be used as input to enhance accuracy, and the learning approach needs to be improved to be applied effectively in the clinical domain. [19] sampled the input by converting the large image into the spectral patches, which is used in the convolutional neural network for image classification.

They evaluated the performance by counting the steps and using the cross-validation method, which is also the benchmark for the proposed solution. This solution has improved the current state of art solution [20]. The author provided the solution of using 37-fold, leave-one-out external-validation which showed the reliability of classification and suggested that it can be used for any new patient. While this solution provides a reliable method of classifying the classifying normal and cancerous tissue from the hyperspectral image with better accuracy but the classification is totally based on the few data samples. Thus this problem can be solved by using the larger data set with more patient HSI data.

[21] has modified the convolutional neural network (CNN) of AlexNet to fc-CNN model by modifying 3 dense layers at the end to operate it on a sliding window. This solution has improved the current state of art solution [22]. It helped to process the arbitrarily large input which uses Region of Interest (ROI) as input. The accuracy gained by Active Learning (AL) has been improved by 3% as compared to Random Learning [23], without requiring any additional cost to training. While the classification accuracy of active learning applied in 3 iterations is increased by 3% over random learning, stain normalization on training has not been considered in the experiment. Thus, the proper normalization of data needs to be done to avoid the generalization errors. [24] improved the current predictive model by considering the tumour depth as a measuring factor. The author applied recursive feature elimination to determine the most important feature to optimize the classifier performance. Which improved the current state of art solution [25]. They offered the solution by developing the classificaiton algorithms to predict pathological lymph node metastasis which maximized the area under the receiver operating characteristic curve (AUC) to 0.840 using decision forest tree algorithm. While the classification performance has been improved to 0.840 as compared to another predictive model, the quality of input data set has not been considered in the experiment. The algorithm needs to consider the quality of data in which the actual depth of invasion (DOI) is measured according to the accepted standard.

[23] used their own Fully Conventional Neural Network (FCN), CN24 and the initialization of the weights was done by pretraining the open-source ILSVRC2012 dataset for improving the training data set. This solution has improved the current state of art solution [26]. They also divided the whole image size to 384*384 region tiles for better processing and reduction in memory. Hence, they used a pixel-wise classification of the imaged tissue, which helped to increase the accuracy in overall recognition rates of 75% and 83%. While the overall recognition rates of the input images have been improved, but there is still doubt if pixels of head and neck cancer and other tissues using multimodal images alone is enough for the classification. The other types of pathological images of the tissues should be considered while conducting the experiment. [27] enhanced the current deep convolutional neural network (DCNN) to improve the average validation accuracy. They modified VGG-16 CNN by removing the first fully-connected layers and adding a new layer with three units to convert the output to a three-class probability. This solution has improved the current state of art solution [28]. They used a softmax layer to convert the output of the fully connected layer and used VGG-16 parameters pretrained with IMAGENET for transfer learning. It provided the best average validation accuracy of 60.7%, 64.7%, and 68.0%, for the image size of 56, 112, and 224, respectively. While the average validation accuracy has been improved for 2D computed Tomography images [5], 3D images need to be considered for the better accuracy for ternary classification. The author has only investigated the effect of small size images on the output, but they need to determine the effect of bigger image size.

[29] designed an adapted convolutional neural network (CNN) architecture to the characteristics of mass spectra. They introduced the analysis tool based on the sensitivity of the input-output relationship to allow interpretation in the input domain. This solution has improved the current state of art solution[30]. It revealed the model artifacts that distinguished lung and primary pancreas tumour. The current solution used the Imaging mass spectrometry (IMS) data but has not considered the pixel-to-pixel variation in data. The source data might trigger some technical issues like misalignment or variation of sample data of different patients. Thus the experiment can include the study of design which involves data from various institutions, patients or operating devices the data should be labelled and aligned with the design choices. [31] has modified the VGGNet to design the

patch-based DCNN architecture with data augmentation process. They combined T2-weighted imaging (T2WI), diffusion-weighted imaging [24], and apparent diffusion coefficient (ADC) and used one high b-value DWI image, which achieved better diagnostic accuracy than using three different b-values. The performance has been improved with the area under curvature value of 0.944 with 95% confidence interval compared to the traditional prediction model. But the author has not considered the data set size and parameter initialization, which need to be considered for better performance.

## 2.1 State of Art Solution

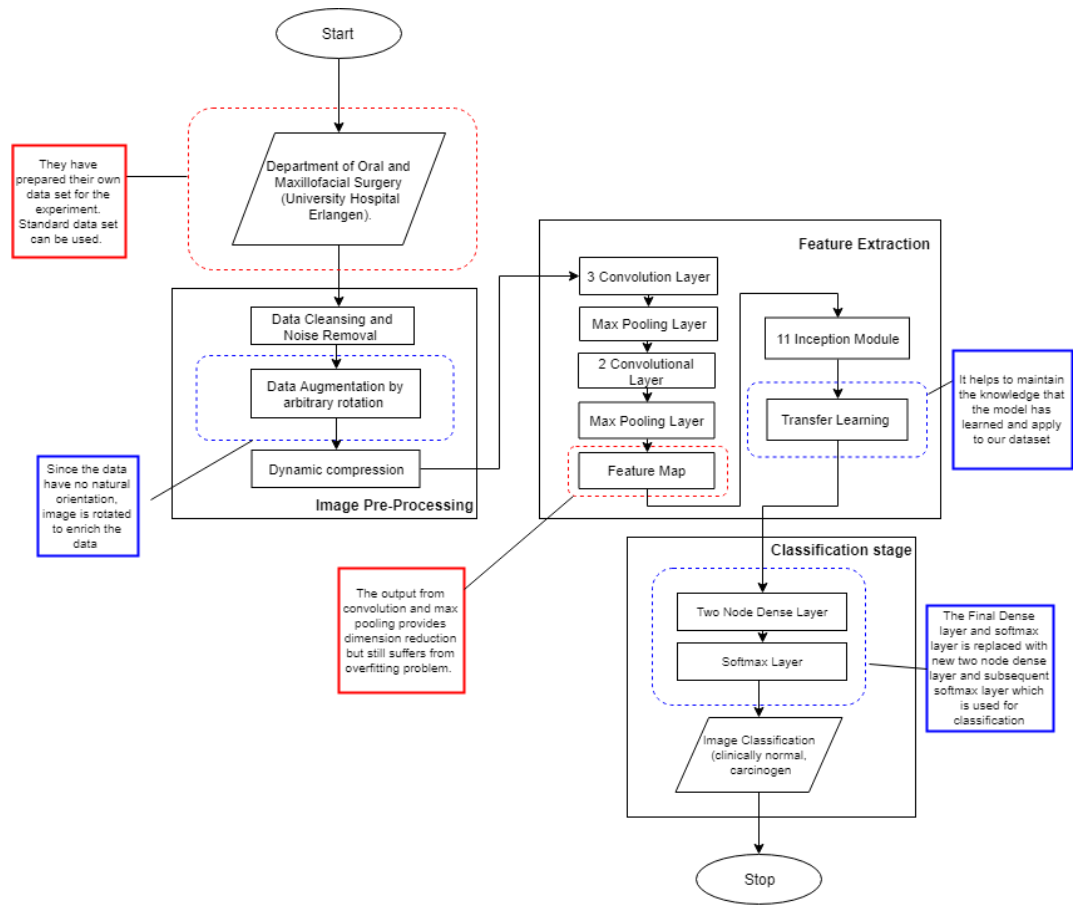

**Fig 1:** Block Diagram of the State of Art System, (Aubreville et al., 2017)
(The blue borders show the good features of this state of art solution, and the red border line shows the limitation).

This part presents the features of the current system (inside the broken blue line in Fig. 1) and limitations (inside the broken red line in **Fig. 1**). [4] used the pre-trained CNN, the Google Net Inception V3 Convolutional Neural Network using ImageNet that were pre-trained on real-world images and is fine-tuned on the new image data set using the transfer learning mechanism. The feature extraction was done with the convolution filters and downsampled by the max-pooling filters which generate the important feature map. The final dense layer and soft-max layer were replaced with a new two-node dense layer, and a subsequent layer was added for image classification [4]. It provides leave-one-patient out cross-validation accuracy of 87.02%, and AUC is 0.948. This model consists of three stages (**Fig 1**: State of Art) i.e. Image Pre-Processing stage, feature extraction stage and classification stage. See figure 2.

*Image Pre-processing:* The sample images are acquired from the Department of Oral and Maxillofacial Surgery (University Hospital Erlangen). The suspected carcinogenic region image was resected after being acquired. Noise is removed by the filtering process, and other good quality images remain for image recognition [4]. Since the image does not have a natural orientation, data augmentation is applied to the classifier by randomly rotating the images. Since Confocal Laser Endomicroscopy (CLE) data is 16 bit, and the Inception v3 model only accepts 8-bit image input, dynamic compression is applied to scale the image.

*Feature extraction:* In this stage, important features are extracted using Google Net Inception V3 Convolutional Neural Network architecture [4]. This architecture consists of two convolution layers and two max pooling layers and nine inception layers with each inception layer consisting of six convolution layers and one pooling layer, as shown in **Fig 1**. This approach of transfer learning helps to decrease the training time and size of the data set required to train the model. Different class images are used for pre-training the network, and the learned knowledge is applied to this dataset. Also, it dramatically reduces the number of parameters in the network, eliminating a large number of irrelevant parameters. This network is trained with 3000 epochs of 100 steps using Adam optimizer with a step size of 0.01 [4]. The input image is squared in the middle of the CLE view area; as a result, information might be lost discarding the 36% of the available area. Due to using simple convolution and max pooling filters at the starting point of the model, it might not exactly produce the required dimension-reduced feature map. Image border might contain information like dysplastic or carcinogenic tissue characteristics, which might be used in mage classification to find out the carcinogenic tumor cell. This information might be discarded due to using rounded images.

*Classification:* From the Google Net Inception V3 CNN architecture, the final dense layer and the softmax layer are replaced with a new two-node dense layer and the subsequent soft-max layer which classify the images as per the training [4]. This model presents the cross-validation accuracy of 87.02% and AUC is 0.948 [4]. The convolution and max pooling filters are used to segment the image, reduce the dimension of the input to avoid the overfitting problem. However, still, accuracy can be increased by solving the overfitting problem, and processing time can be reduced.

The neuron is the part of the hidden convolution layer and out energy of the neuron $E_{j,k}$ is given by:

$$E_{j,k} = f(b + \sum_{i=0}^{q} \sum_{z=0}^{l} W_{i,z} * X_{i,z}) \quad (1)$$

The energy output E is calculated for neuron k in layer j of the CNN where we use $f$ (ReLU activation function) to calculate the fire value.

'b' is the bias factor
'W' is the weight of each neuron in layer j-1 for the input connection
'X' is the input from the nodes in the previous layer with convolution
'q' and 'l' represents the size of the input matrix of shared weights of W
'ii' and 'jj' are the indexes of the input activation at position (j+i, k+z)

**Table 1**: **Convolution and max pooling method algorithm**

| Algorithm: Convolution and max pooling method to reduce the input dimension |
|---|
| Input: CLE image |
| Output: Feature map without appropriate input dimension reduction |
| BEGIN |
| Step 1: Get the number of iteration (l) for each image CLE input image |
| Step 2: Get the number of iteration (q) for each image row and set the value of accumulator to zero |
| Step 3: The input is multiplied with respective weights as w1*x1 + w2*x2 + w3*x3+…+wn*xn and stored on the accumulator and move for the next layer |
| Step 4: End loop of the layer of step 2 and go to step 1 and increase the value of l by 1, i.e. next neuron |
| Step 5: end loop 1 |
| Step 6: Each perceptron also has a bias which can be thought of as how much flexible the perceptron is. It is somehow similar to the constant b of a linear function y = ax + b. It allows us to move the lineup and down to fit the prediction with the data better. |
| Step 7: Add Step 7 bias value with the Step 3 output result and store the final result in the accumulator |
| Step 8: Apply the $f$ (ReLU activation function) to step 5 to get the energy output E |
| END |

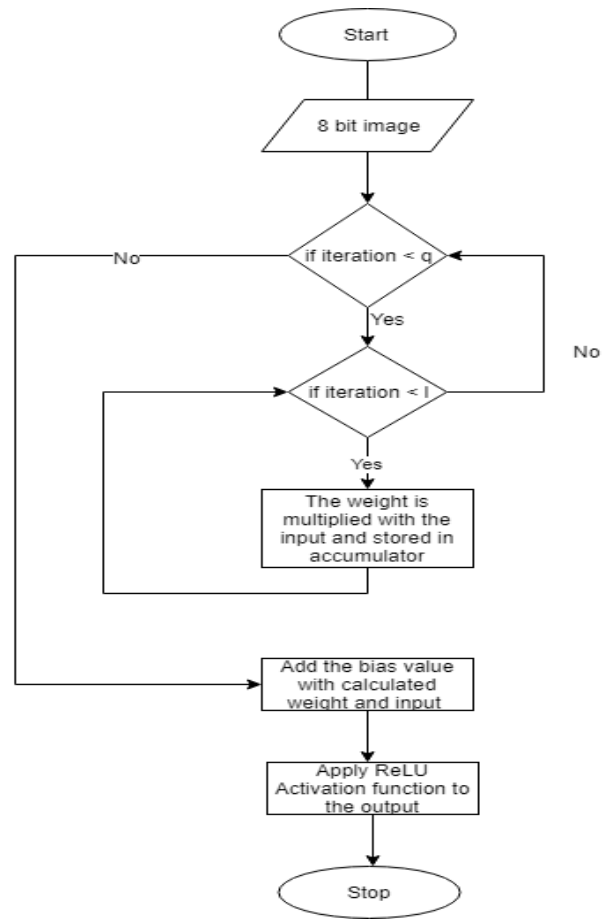

**Fig 2:** Flow chart of Convolution and Max Pooling algorithm

## 3. Proposed System

Oral cancer image classification techniques' review has led to the analysis of the pros and cons of each method. The main issues of those systems were errors in classification, longer processing time and overfitting; which need to be addressed. From our review, we selected the best one, [4] as the basis for our proposed solution from the Inception v3 transfer learning method for better feature map and accurate classification. Inception v3 transfer learning retrained the final layer of an existing model, which significantly decreased training time, and dataset size, which is required for training the model. Thus, it maintains the learned knowledge from the training and can be applied to the desired data set, resulting in highly accurate classifications without need for extensive training and computational power. Moreover, the proposed solution with autoencoder implementation [5], extracts characteristic features from given inputs by learning a network, which reproduces input data from those features. The input data are scanned by the convolution filter and downsampled by a max pooling layer and is passed to the encoding layer where deconvolution and depooling is done to extract the optimised image input. This helps to extract more complex image features with higher-order structures. The proposed system consists of three main stages (**Fig. 3**) – Image Pre-processing, Feature Extraction and Classification. See figure 4 and table 2.

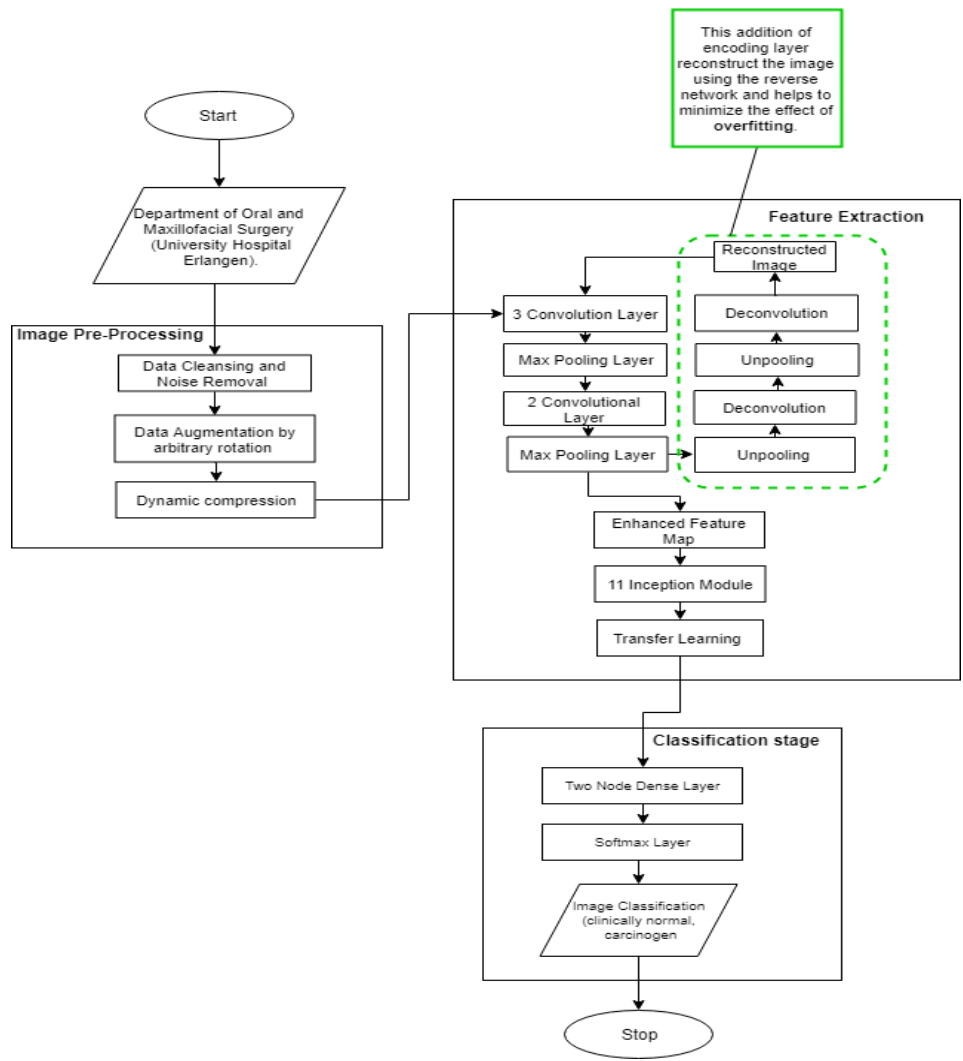

**Fig 3:** Block Diagram of the proposed deep learning system for oral tumor classification using autoencoder architecture
[The green borders refer to the new parts in our proposed system]

*Image Pre-processing:* The suspected carcinogenic regions were resected after image acquisition. Noise removal or filtering is used to enhance the quality of the images, which are subsequently used for image recognition [4]. The two-fold data augmentation is applied to the image to enrich the data provided to the classifier by random rotation of the image. Since CLE data is 16 bit and the Inception v3 model only accepts 8-bit image input, dynamic compression is applied to scale the image.

*Feature extraction:* In this stage, the Google Net Inception V3 Convolutional Neural Network architecture is applied to extract the important features [4]. First, the input data are scanned by convolutional filters; then the image is down-sampled by a max-pooling layer, which is finally passed to encoding layer, where un-pooling and de-convolution are done to generate the input data to minimize the difference between input and output data (Antonio et al., 2018) as shown in **Fig 3**. The sparsity penalty function is introduced to enhance the efficiency of the feature extraction and to compress the information in the autoencoder as shown in **Fig 3**. Essential features like lesion patch size and microscopic features are extracted as a feature map for the oral cancer classification. The overfitting problem is minimized by adding the penalty. The transfer learning helps decrease the training time and size of the training data set. The learned knowledge from the pre-training is applied to this dataset. Also, it dramatically reduces the number of parameters in the network, eliminating a large number of irrelevant parameters.

*Classification:* From the Google Net Inception V3 CNN architecture, the final dense layer and the softmax layer are replaced with a new two-node dense layer and subsequent softmax layer which classify the images as per the training [4]. The different convolutional and pooling layers are usually used to extract more abstract feature representations in moving through the network. This two nodes dense layer and subsequent softmax layer classify the output of the neural network. The output is the probability distribution of all classes, which uses loss function to measure the difference between the actual output and the target output.

Enhancing feature extraction efficiency and compressing the information in auto-encoder has increased the accuracy by adding the sparse penalty, as shown in **equation 2**.

$$X_{ii,jj} = R + \lambda_s S \tag{2}$$

Where,
$\lambda_s$ is a weight constant
R is the optimisation factor
S is the sparsity constant

For features extraction , Eq 3equation 4:

$$S = 1^n \sum_{j=1}^{l}(-r_j^{encode} \log r_j^{encode}) \tag{3}$$

Here, $r_j^{encode}$ is the output intensity of filter j in the encoding layer relative to their total summation.
x is the input
q and l are the number of nodes in the input and encoding layers.
$\lambda_s$ is a weight constant

The total network can be optimized by reducing the difference between the input and the output by using the reversed network in which the information is also encoded to enhance the feature extraction, as shown in **equation 4**. The features can be extracted during the process of optimising the input parameters by reducing the reconstruction error [32], which is given by the cost function as:

$$R = 1/n \sum_{i=1}^{n} 1/2(|x(i) - y(i)|^2) + \lambda/2(||W||)^2 \tag{4}$$

$x$ is the input
y is the reconstruction of x
W is the weight matrix
$\lambda$ is the weight decay parameter
i, j are the hidden layer
n is the data size

Equation 4 is now modified by Equation 5 based on the cost function which is given by the mathematical formula as:

$$R' = 1/n \sum_{i=1}^{n} 1/2 + \lambda/2(||W||)^2 \tag{5}$$

Where,

W is the weight matrix
$\lambda$ is the weight decay parameter
i, j are the hidden layer
n is the data size

The reconstruction error calculated as shown in equation 6

Prakrit Joshi, Omar Hisham Alsadoon, Abeer Alsadoon, Nada AlSallami, Tarik A. Rashid, P.W.C. Prasad, Sami Haddad (2022). Deep learning for size and microscope feature extraction and classification in Oral Cancer: enhanced convolution neural network. *Multimed Tools Appl*. https://doi.org/10.1007/s11042-022-13412-y
$$MR = 1/n \sum_{i=1}^{n} 1/2(x_i^{output} - x_i^{input})^2 + \lambda/2(||W||)^2 \tag{6}$$

Where,
W is the weight matrix
λ is the weight decay parameter
$x$ is the input
y is the reconstruction of x

Here, the squared-error cost function is considered to decrease the magnitude of the weights and prevents from overfitting issue.

sparsity constrain to reduce the reconstruction error given as equation 7

$$MX_{ii,jj} = MR + \lambda s \, S \tag{7}$$

$MX_{ii,jj}$ is the modified input from the nodes
MR is the modified reconstruction error value
*λs* is a weight constant
S is stacked autoencoders

Finally, equation 1 [4] has been enhanced by us by modifying sparsity constrain suggested by [32] which is expressed mathematically below as equation 9:

$$EE_{j,k} = f(b + \sum_{i=0}^{q} \sum_{z=0}^{l} W_{i,z} * MX_{ii,jj}) \tag{8}$$

Where,
'b' is the bias factor
'W' is the weight of each neuron in layer j-1 for the input connection
$MX_{ii,jj}$ is the modified input from the nodes
'q' and 'l' represents the size of the input matrix of shared weights of W,
'ii' and 'jj' are the indexes of the input activation at position (j+i, k+z).

### 3.1 Area of Improvement

We proposed one equation. The accuracy value of the model is comparatively low due to extracting irrelevant characteristics. The feature extraction efficiency is enhanced, and the information in autoencoder is compressed by introducing the sparsity penalty, as shown in equation (2). This helps in minimizing overfitting of the network, but still, reconstruction error is generated due to the iteration in training data which is shown in equation (3). The reconstruction error is reduced by modifying equation 3 with the help of equation (5). The difference between the input and the output in the network can be reduced by modifying equation (5) with the help of equation (6) to obtain the final modified reconstruction error equation (7). This equation (7) aim is to decrease the magnitude of the weights that helps prevent overfitting where weight decay parameter λ is introduced. Equation (2) is modified to obtain the equation (8) which is modified sparsity constrain to reduce the reconstruction error. Hence, with the help of equation (8), the feature extraction process is enhanced as shown by equation (9) to generate the final feature map, which improves the overall system classification accuracy.

### 3.2 Why Autoencoder with the reverse network?

With autoencoder used along with the inception V3 layers, the number of features is reduced to produce reasonable features. This proposed system adds an encoding layer in which the output from convolution and max pooling layer is passed to the reversed network, where unpooling and deconvolution are done to generate the input data to minimize the difference between input and output data for reconstructing the image. The feature extraction process is enhanced, and the information is compressed by the introduction of sparsity penalty, which minimizes the effect of overfitting in the network. Reducing the overfitting problem in the network helps to enhance the efficiency of the feature extraction and generate the exact important feature map by training the network, which is used for the classification purpose. Thus such features help to improve the classification accuracy by providing exact information for classification, which is also captured by transfer learning function to reduce the processing time. See Table 1 and Fig 3. As seen from the literature in this paper, the available solutions have directly fed the complete image to the convolution and max pooling layers which is then passed to inception V3 layers. None of the solutions has considered the network complexity and reason behind increasing the number of parameters and considering the correlated variables. Our proposed solution has reduced the number of parameters in the generated feature map using autoencoder. It increases the classification accuracy by information compression and reconstruction of images.

Table 2: Proposed autoencoder with deconvolution and depooling method to reduce the input dimension

| |
|---|
| Algorithm: Proposed autoencoder with deconvolution and depooling method |
| Input: CLE image of oral tumor |
| Output: Feature map without appropriate input dimension reduction |
| BEGIN |
| Step 1: Get the number of iteration (l) for each image CLE input image |
| Step 2: Get the number of iteration (q) for each image row and set the value of accumulator to zero |
| Step 3: The input is multiplied with respective weights as w1*x1 + w2*x2 + w3*x3+…+wn*xn and stored on the accumulator and move for the next layer |
| Step 4: Pass the output to the encoder layer where the input is reconstructed and information is compressed and use that reconstructed image as an input and repeat step 3. |
| Step 5: End loop of layer of step 2 and go to step 1 and increase the value of l by 1, i.e. next neuron |
| Step 6: end loop 1 |
| Step 7: Each perceptron also has a bias which can be thought of as how much flexible the perceptron is. It is somehow similar to the constant b of a linear function y = ax + b. It allows us to move the line up and down to fit the prediction with the data better. |
| Step 8: Add Step 7 bias value with the Step 3 output result and store the final result in the accumulator |
| Step 9: Apply the $f$ (ReLU activation function) to step 5 to get the energy output E |
| END |

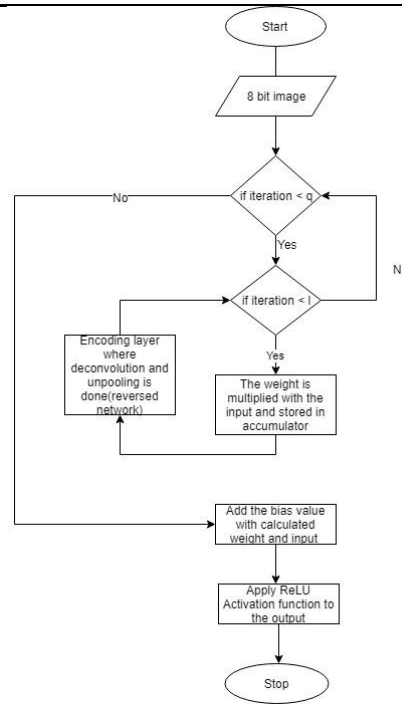

**Fig. 4:** Flow chart of Convolution and Max Pooling with autoencoding

## 4. Results and Discussion

Python 3.6.0 with Sliderunner were used for the implementation of the model using 32 different CLE images samples from different datasets of varying age groups. Sliderunner was used to cell annotations in whole slide images. Two convlutuonal layers were use the first convulational with 64 filters of (5x5)px, followed by (3x3) px max pooling layer. The second convolutional layer with 32 filters followed by (3x3)px max pooling layer.

The sample data consist of images from various stages of radiotherapy treatment i.e. pre-treatment, mid-treatment, and post-treatment. The data samples exclude the participant such as pregnant women, patients younger than 18 years and patients whose size and weight would not allow scanning. The sample dataset used in this training is a free open source, just as the folder that can be downloaded from the cancer imaging archive. The CNN with inception module is used for the implementation based on a deep learning algorithm. There are three groups of sample images taken from the three different anatomical locations such as vocal cord, oral cavity and samples from both vocal fold and oral cord area. Those three different anatomical images have been tested for the classification accuracy as shown in Tables 3, 4 and 5. We used 1.6 GHz Octa Core i5 processor with 8 GB RAM memory for the experiment. The image quality is at an accepted level, and we configured our server and data storage to support our purpose.

The patch from the whole image is extracted by using the file Patch extraction (and randomization) (extractPatches.py) in python. The file cellVizio is read, and the patches without annotated artifacts are extracted as per the value entered in the CLE database. The microscopic feature to determine the presence of lesion is extracted from the CLE image to classify it as a carcinogenic image of oral cancer. The classifier was trained and tested using the train.py file in which the numpy array with images probability is generated. We used 33% of the dataset for testing the network using 10 fold cross-validation. The patch probability calculation is done by equation 10 in Result Analysis table; Finally, the classification is done based on the patch probability function, which is based on the areas of the image that are covered multiple times by patches. The library used to calculate the confidence range for Python was Numpy. The accuracy is calculated using the confidence range shown by equation 11. During the feature extraction stage, the modified CNN extracts the important features from the image input based on the labelled data, which helps exact classification. See figure 5.

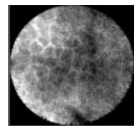

Fig. 5: Input image after passing to Convolutional Neural Network and max pooling layer

The output from the convolutional layer and the max pooling layer is again passed to the autoencoder where the de-convolution and un-pooling are done to reconstruct the image and encode the information which is again passed to the same convolutional layer to generate the better feature maps as shown in the **figure 6** below:

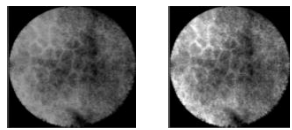

Fig. 6: Image reconstruction after passing through reverse network for un-pooling and de-convolution

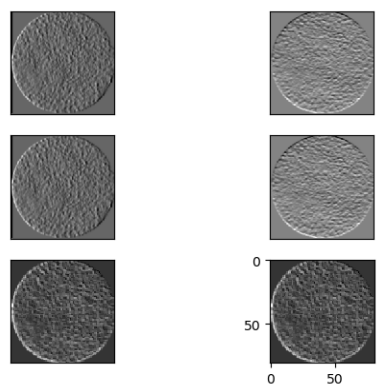

Fig. 7: Extracted features of input reconstructed image passed from the autoencoder

In Fig 7, feature map is generated by inputting the reconstructed and encoded image from the convolution and maxpooling layer. Usually the square patches are generated from the round field of a Confocal Laser Endomicroscopy (CLE) image to generate important feature map. The extracted feature is then passed to the inception layers for better learning of the network.

Finally, the selected features are used for image classification. Based on the training and testing of the CLE image data sets, the sample images that have been classified is shown in table 3, 4 and 5. See figures 8 and 9.

Graphs and tables were used for the comparison of image sample between the proposed solution and state of art solution. The results obtained after the classification of the CLE images from the different anatomical location of squamous cell carcinoma are evaluated in table 3, 4 and 5. The results are tabulated in the table according to the different group of CLE images. And the results from the test samples are represented in terms of accuracy and processing time. The probability score labelled after the classification is used to measure accuracy. The test is carried out with the three different subjects of sample image data. Based on the microscopic feature i.e. feature of dysplasia, each test images of different anatomical locations is classified either as clinically normal or verified carcinoma i.e. presence of a lesion. The average accuracy and average processing time are measured for each group classified.

The result is compared to the classification stage of the system. Deep learning is used for image classification. The proposed solution has enhanced the accuracy of the classification by employing autoencoder in the feature selection process and also has reduced the processing time by avoiding the random generation of the input weight matrix in the classification process. This will improve in the early prediction of oral cancer.

Table 3: Accuracy and Processing time results for squamous cells cancer classification (Sample image Dataset 1 with **microscopic feature** from **HNSCC-3DCT-RT**)

| Sample No. | Sample group details | Original Images | State of Art | | | Proposed solution | | |
|---|---|---|---|---|---|---|---|---|
| | | | Processed sample | Accuracy (%) | Processing time (sec) | Processed sample | Accuracy (%) | Processing time (sec) |
| 1.1 | **Vocal fold area** | 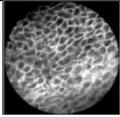 | 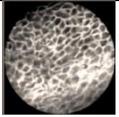 | 88.21% | 0.399s | 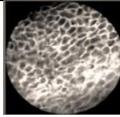 | 93.02% | 0.412s |
| 1.2 | | 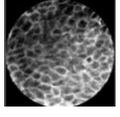 | 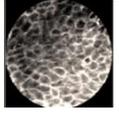 | 87.85% | 0.403s | 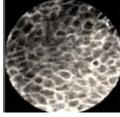 | 92.45% | 0.389s |
| 1.3 | | 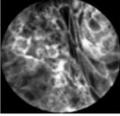 | 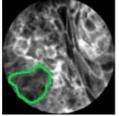 | 87.16% | 0.412s | 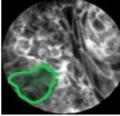 | 92.77% | 0.387s |
| 1.4 | | 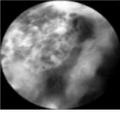 | 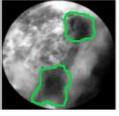 | 85.78% | 0.390s | 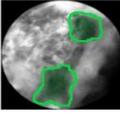 | 91.67% | 0.331s |
| 1.5 | **Oral Cavity** | 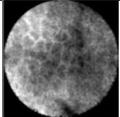 | 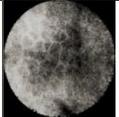 | 86.67% | 0.375s | 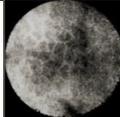 | 91.23% | 0.386s |
| 1.6 | | 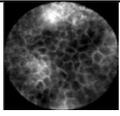 | 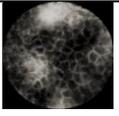 | 86.21% | 0.366s | 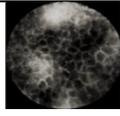 | 91.98% | 0.345s |
| 1.7 | | 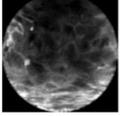 | 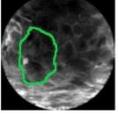 | 85.98% | 0.347s | 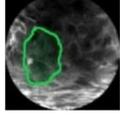 | 90.98% | 0.323s |
| 1.8 | | 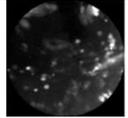 | 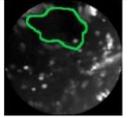 | 86.89% | 0.392s | 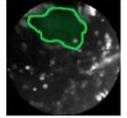 | 92.01% | 0.349s |
| 1.9 | **Vocal Chord+ Oral Cavity** | 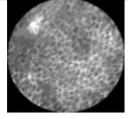 | 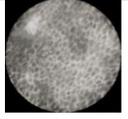 | 87.32% | 0.386s | 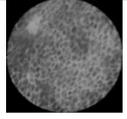 | 92.67% | 0.315s |
| 1.10 | | 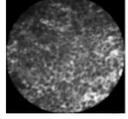 | 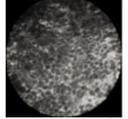 | 86.54% | 0.372s | 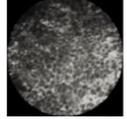 | 91.98% | 0.375s |
| 1.11 | | 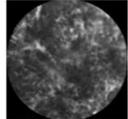 | 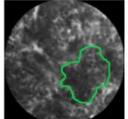 | 87.05% | 0.397s | 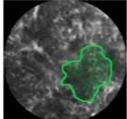 | 90.89% | 0.312s |
| 1.12 | | 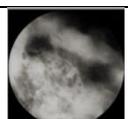 | 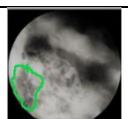 | 85.89% | 0.356s | 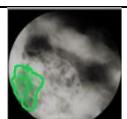 | 91.67% | 0.335s |

Table 4: Accuracy and Processing time results for squamous cells cancer classification (Sample image Dataset 2 with identified image **patch size** study taken from **TCGA-HNSC**)

| Sample No. | Sample group details | Original Images | State of Art | | | Proposed solution | | |
|---|---|---|---|---|---|---|---|---|
| | | | Processed sample | Accuracy (%) | Processing time (sec) | Processed sample | Accuracy (%) | Processing time (sec) |
| 2.1 | **Vocal fold area** | 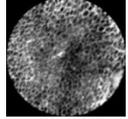 | 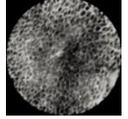 | 87.11% | 0.409s | 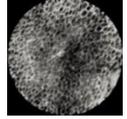 | 91.63% | 0.391s |
| 2.2 | | 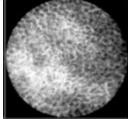 | 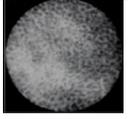 | 86.16% | 0.373s | 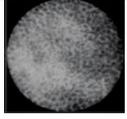 | 92.67% | 0.328s |
| 2.3 | | 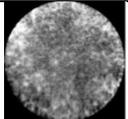 | 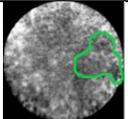 | 86.03% | 0.384s | 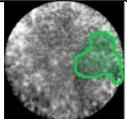 | 91.62% | 0.341s |
| 2.4 | | 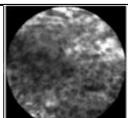 | 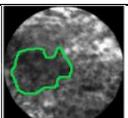 | 87.34% | 0.456s | 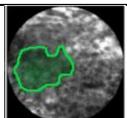 | 91.89% | 0.354s |
| 2.5 | **Oral Cavity** | 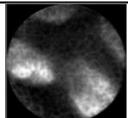 | 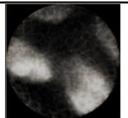 | 85.98% | 0.352s | 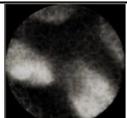 | 91.13% | 0.315s |
| 2.6 | | 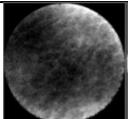 | 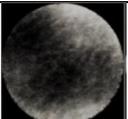 | 86.33% | 0.377s | 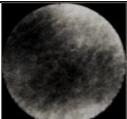 | 91.34% | 0.307s |

| Sample No. | Sample group details | Original Images | Processed sample | Accuracy (%) | Processing time (sec) | Processed sample | Accuracy (%) | Processing time (sec) |
|---|---|---|---|---|---|---|---|---|
| 2.7 | | 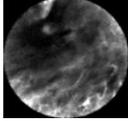 | 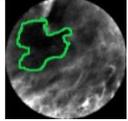 | 85.62% | 0.382s | 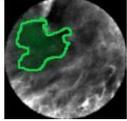 | 91.17% | 0.327s |
| 2.8 | | 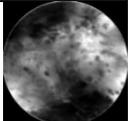 | 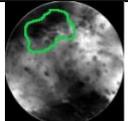 | 85.86% | 0.381s | 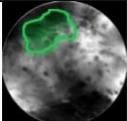 | 90.89% | 0.342s |
| 2.9 | **Vocal Chord+ Oral Cavity** | 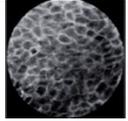 | 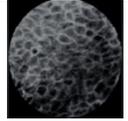 | 86.77% | 0.349s | 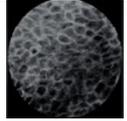 | 90.23% | 0.291s |
| 2.10 | | 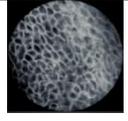 | 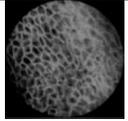 | 85.89% | 0.356s | 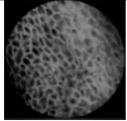 | 91.67% | 0.335s |
| 2.11 | | 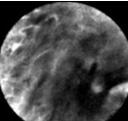 | 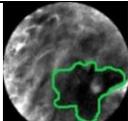 | 84.79% | 0.397s | 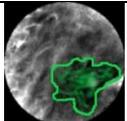 | 91.66% | 0.331s |
| 2.12 | | 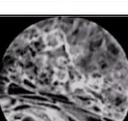 | 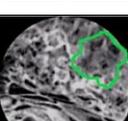 | 84.95% | 0.365s | 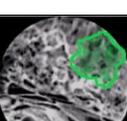 | 90.89% | 0.354s |

Table 5: Accuracy and Processing time results for squamous cells cancer classification (Sample image Dataset 3; **BIO-Imaging**)

| Sample No. | Sample group details | Original Images | State of Art | | | Proposed solution | | |
|---|---|---|---|---|---|---|---|---|
| | | | Processed sample | Accuracy (%) | Processing time (sec) | Processed sample | Accuracy (%) | Processing time (sec) |
| 3.1 | **Vocal fold area** | 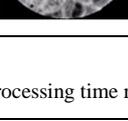 | 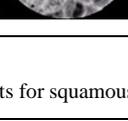 | 88.67% | 0.432s | 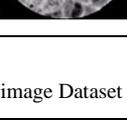 | 93.12% | 0.411s |
| 3.2 | | 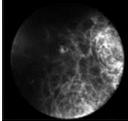 | 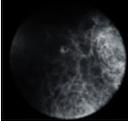 | 87.67% | 0.412s | 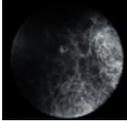 | 92.12% | 0.371s |
| 3.3 | | 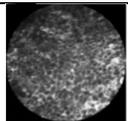 | 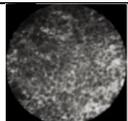 | 88.17% | 0.393s | 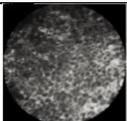 | 92.82% | 0.361s |
| 3.4 | **Oral Cavity** | 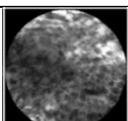 | 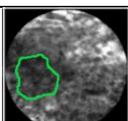 | 87.41% | 0.371s | 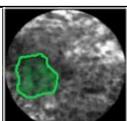 | 92.34% | 0.319s |
| 3.5 | | 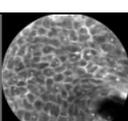 | 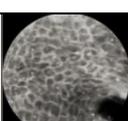 | 87.33% | 0.357s | 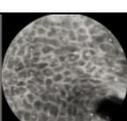 | 91.11% | 0.331s |
| 3.6 | | 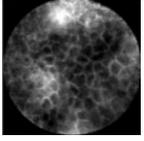 | 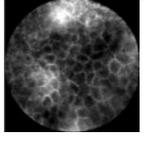 | 86.89% | 0.378s | 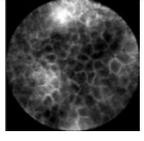 | 91.99% | 0.315s |
| 3.7 | **Vocal Chord+ Oral Cavity** | 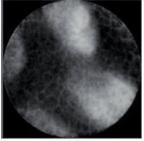 | 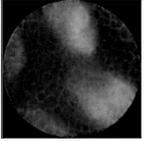 | 87.89% | 0.356s | 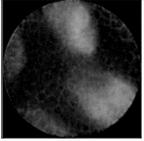 | 92.67% | 0.335s |
| 3.8 | | 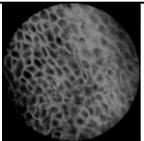 | 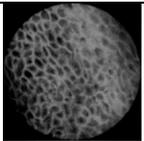 | 85.97% | 0.345s | 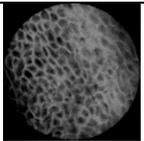 | 91.12% | 0.297s |
| 3.9 | | 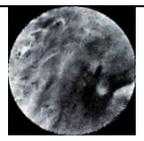 | 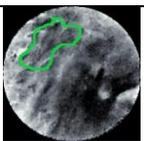 | 86.67% | 0.326s | 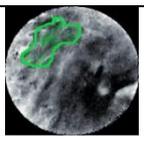 | 91.61% | 0.301s |

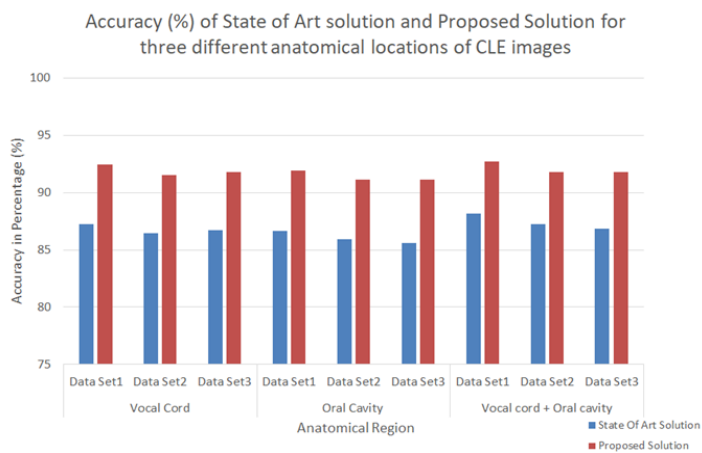

Fig. 8: Bar graph shows the processing in percentage for three different anatomical locations of squamous cells of CLE images. The blue color indicates the accuracy of State of Art solution while orange colour indicates the accuracy of the proposed solution. (1) First couple, second couple and third couple of bar graph shows the average accuracy for vocal cord region for two different data sets (2) fourth couple, fifth couple and sixth of bar graph shows the average accuracy for oral cavity region for two different data sets (3) seventh couple, eight couple and ninth of bar graph shows the average accuracy for both vocal cord and oral cavity region for three different data sets.

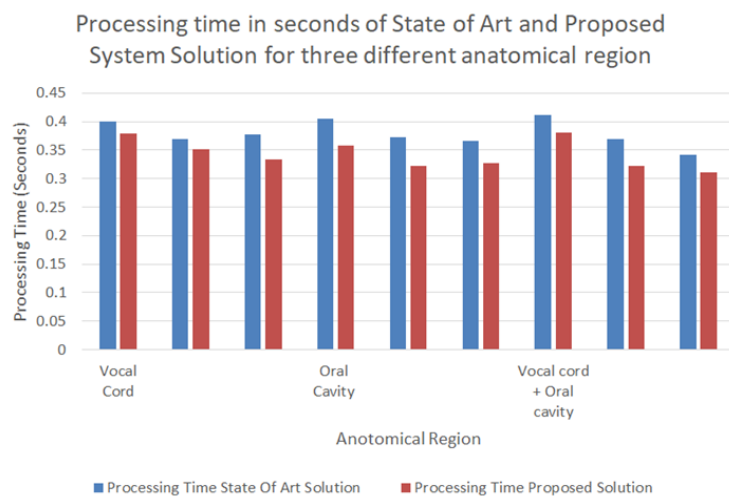

Fig. 9: Bar graph shows the processing time in seconds for three different anatomical locations of squamous cells of CLE images. The blue color indicates the accuracy of State of Art solution while orange color indicates the accuracy of proposed solution. (1) The first couple,

second couple and the third couple of bar graph shows the average accuracy for vocal cord region for two different data sets (2) fourth couple, fifth couple and sixth of bar graph shows the average accuracy for oral cavity region for two different data sets (3) seventh couple, eight couple and ninth of bar graph shows the average accuracy for both vocal cord and oral cavity region for three different data sets

The results illustrate that there is an overall improvement in accuracy and processing time as compared to the state of art solution for the classification of images. The proposed system enhances the classification accuracy to 92% with the help of modified Convolutional Neural Network which is better than state of art solution. It processes the reconstructed image which requires less training than the original image to learn the feature. Similarly, processing time decreases by 20 to 30 milliseconds with the help of transfer learning mechanism that records the previous processing learning and fine-tunes the new image data set through the learning. The patch probability score is used to measure the accuracy whereas execution time is used to calculate the processing time of the system. Processing time is measured in second and accuracy is measured in percentage.

The current Convolutional Neural network using inception V3 module layer is modified using autoencoder which is used in feature extraction stage that enhances the performance of the proposed system. Python programming language was used for the implementation, which has minimized the processing time and increased the classification accuracy of the model. The use of autoencoder, reconstruction and enhancement of the image and encode the information which is then again used as an input for extracting significant features. The purpose of the two dense layer and softmax layers in a convolutional neural network for the classification has minimized the processing time by avoiding the random generation of the weight matrix. In conclusion, the Convolutional Neural Network combined with the autoencoder and dense layer has enhanced the CLE image classification with an overall improvement of accuracy by 5% and processing time by 20 ~ 30 milliseconds.

There are various techniques and algorithms that have been implemented for classifying CLE images. However, many alternatives and techniques have been researched to improve the accuracy and processing time of the classification. The limitation of the current best solution has been solved in this research with improved accuracy of 92% against the current accuracy of 87%. The system has reduced the processing time to 0.343 seconds from 0.380 seconds. This improvement in accuracy and processing time is a result of implementing the autoencoder in the feature extraction process to generate the important feature map hence solving the overfitting issue. The proposed system has better accuracy and reduced processing time in all different sample image groups of oral section such as oral cavity, vocal cord and both regions. See Table 6.

Table 6: Comparison table between Proposed Solution and State of Art Solution

|  | **Proposed Solution** | **State of Art Solution** |
|---|---|---|
| **Name of the solution** | Inception V3 with Autoencoders | Inception V3 CNN |
| **Accuracy** | Improved the feature map by reducing the input dimension. Provides an accuracy improvement in classification of 92% | Provides an accuracy of 87% |
| **Processing Time** | Decrease in processing time from 0.380 milliseconds to 0.343 milliseconds | Provides a processing time of 380 milliseconds |
| **Proposed equation** | $EE_{j,k} = f(b + \sum_{i=0}^{q} \sum_{z=0}^{l} W_{i,z} * MX_{ii,jj})$ | $E_{j,k} = f(b + \sum_{i=0}^{q} \sum_{z=0}^{l} W_{i,z} * X_{ii,jj})$ |
| **Contribution 1** | Autoencoder helps to minimize the different between the input and the output by sending the output from max pooling to the reverse network which help to optimize the total network | The state of art does not consider the optimisation of inputs and calculates the energy as the original value. |
| **Contribution 2** | It helps to encode the information by introducing sparsity penalty for the optimization function. | The state of art select only one variable from a group of highly correlated variables and ignores others. |

# 6. Conclusion and Future Work

The accurate classification of oral cancer tumour in confocal laser endomicroscopy (CLE) images is essential for the early diagnosis of oral cancer. Deep learning technology has been successfully implemented for the identification and classification of oral squamous cell carcinoma. The purpose of this paper is to improve the accuracy and to reduce the processing time of squamous cell carcinoma classification. The autoencoder has been developed by combining it with the Inception V3 module, which is adopted in the second-best solution (Antonio et al., 2018). It adds a sparsity penalty function that helps to minimize the overfitting issue. This solution improves the feature extraction process by reducing the number of parameters in the network by reconstructing the image after passing to reverse module. Hence, it enhances the classification accuracy by 5% on average and reduced the processing time by 20~30 milliseconds on average. For future work, a large amount of CLE image dataset can be used for more feature extraction from CNN network to train the module effectively. In future work, a large amount of CLE image dataset can be used for training and testing process of CNN. The network can even be run with the multi-variate dataset with 3D visualisation to produce the exact classification result.

**Appendix:**

| | |
|---|---|
| **RISA** | Reconstruction independent subspace anal-ysis (RISA). |
| **CNN** | Convolutional Neural Network |
| **FCN** | Fully convolutional neural networks |
| **VOI** | Volume of Interest |
| **LBP-TOP** | Rotation invariant uniform pattern local binary pattern on three orthogonal planes |
| **ReLU** | Activations on the convolutional layers are rectified linear units |
| **fc-CNN** | Fully convolutional CNN |

**Compliance with Ethical Standards:**

Funding: No Funding has used in this work.
Conflict of Interest: No conflict of interest